\documentclass[final,5p,times,twocolumn]{elsarticle}

\usepackage{graphicx}
\usepackage{amssymb}
\biboptions{super}
\usepackage{url}
\newfont{\mm}{msbm10}

\journal{Computer Physics Communications}

\begin{document}

\begin{frontmatter}



%\title{Who needs `error propagation laws?'}
%\title{Incredibly simple indirect measurements}
%\title{Interval uncertainty management}
%\title{Reliable uncertainty management using interval calculations}
\title{Reliable uncertainties in indirect measurements}

%version 4, 2017-07-24
%version 5, 2017-07-27
%version 6, 2017-07-28
%version 7, 2017-07-31
%version 7, 2017-08-01


\author{Marek W. Gutowski}

\address{Institute of Physics, Polish Academy of Sciences, 02-668 Warszawa, Poland}
\ead[]{marek.gutowski@ifpan.edu.pl}

\begin{abstract}
In this article we present very intuitive, easy to follow, yet mathematically rigorous,
approach to the so called data fitting process.  Rather than minimizing the distance
between measured and simulated data points, we prefer to find such an area in 
searched parameters' space that generates simulated curve crossing as many
acquired experimental points as possible, but at least half of them.
Such a task is pretty easy to attack with interval calculations.
The problem is, however, that interval calculations operate on guaranteed
intervals, that is on pairs of numbers determining minimal and maximal
values of measured quantity while in vast majority of cases our measured
quantities are expressed rather as a pair of two other numbers: the average
value and its standard deviation.
Here we propose the combination of interval calculus with basic notions
from probability and statistics.  This approach makes possible to obtain
the results in familiar form as reliable values of searched parameters,
their standard deviations, and their correlations as well.
There are no assumptions concerning the probability density
distributions of experimental values besides the obvious one that their
variances are finite.  Neither the symmetry of uncertainties of experimental
distributions is required (assumed) nor those uncertainties have to be `small.'\
As a side effect, outliers are quietly and safely ignored, even if numerous.
\end{abstract}


\end{frontmatter}

% \linenumbers


\section{Introduction}\label{intro}
Practically all known data fitting procedures are based on minimization
of distance between measured and simulated values.  Yet there exist
various, equally good, distances (metrics) in $n$-dimensional
space $\mbox{\mm R}^n$.\  Besides the best known Euclidean
distance there exist many other ones, like Manhattan (taxi driver)
or  Chebyshev distance.  Those three produce exactly the same values
of\  $d(x,y)$\  when $n=1$, but still another  metrics, defined as\
$d(x,y)=c\log\,(1+\left|x-y\right|/c)$\  (with any fixed $c>0$) will differ.

The choice of this or other metrics is therefore not unique.  Exactly
for this reason we have various families of fitting procedures\cite{atmos}
at our disposal, with (weighted) least squares regression (LSQ)  being
most popular.

However, the real question is: why at all we are resorting
to distance--based routines?  The results of individual measurements
are conventionally reported in form $y_{0}\pm\sigma_{y}$, with $y_{0}$
being the most likely numerical estimate of true value $\hat{y}$, and
$\sigma_{y}$ its standard deviation.\  Such result is usually drawn as
a~section of a~straight line
$\left[\,y_{0}-\sigma_{y}\,,\,y_{0}+\sigma_{y}\,\right]$, when presented
in graphs.  This doesn't mean that the true value $\hat{y}=y_{0}$,
nor that $\hat{y}$ is even contained within the interval shown!  Indeed, when
$\hat{y}$ is normally distributed (a~common but often unfounded\cite{notnormal}
assumption), the chance for its true value to be located  outside this
interval is roughly equal to $1/3$ --- rather far from being negligible.\
This simple observation makes all distance--based procedures
questionable at least.\  Besides, we often fit our data after non-linear
transformation, performed just to visualize them as forming a~straight line
in the new coordinate system.\  After such a~transformation, the original
center of uncertainty interval no longer corresponds to the center of its
image, thus making the notion of distance even more dubious.

So, perhaps we should deal with intervals of type 
$\left[\,y_0-\delta,\,y_0+\delta\,\right]$~\
instead, where $\delta$ is maximal uncertainty of a~measurement?
This way the true value $\hat{y}$ is guaranteed to reside inside the
interval shown.  But is it indeed?  No, at least not in experimental practice,
when unpredictable interferences do happen (or just data storage/transmission
errors).\  Additionally, under this approach, our task would be to find a~set
of curves passing through \emph{all} the measured values expressed as
guaranteed intervals. Informally speaking, we should find a~`thick' simulated curve
and then evaluate somehow the uncertainties of all its fitted parameters.
However, it is easy to see that a~single outlier may make this task impossible.
On the other hand, when the said outlier is `small' enough, then the fitted curve
will be unrealistically narrow, thus suggesting incredibly good precision of its
parameters.  No wonder that this approach didn't gain much popularity
in scientific laboratory practice (notable exceptions are proposals given by
Zhilin\cite{EmpDataFit,Zhilin} or Kieffer\cite{ch7}), nevertheless it is developed
in the field of various engineering applications under the keywords like tolerance
problem\cite{Remote,Tolerable}, confidence regions\cite{FIR,B09},
or guaranteed state estimation\cite{Jaul,KieffWalt}.

In many cases the LSQ approach leads to solving a~set of linear equations.\
This is also true for interval versions of those procedures.  It appears, however,
that interval linear systems have definitely more specific features than their
classical predecessors.  Specifically, they may have solutions of many different
kinds.\  This fact was most likely first
recognized by Shary, who also proposed a~classification scheme for solutions
of interval systems of linear equations (applicable to non-linear systems as well).
The literature concerning linear interval equations is now quite
impressive\cite{EmpDataFit,Remote,Tolerable,Jaul,veryOldShary,MCS,
middleShary,Irene,Skalna,Kolev-lin,Kolev1,SergeyIrene,inner,Kolev}
while non-linear systems are investigated less frequently\cite{ch7,FIR,KieffWalt,
Kolev-non,Kolev-nonlin} and often presented at conferences only\cite{Tucker}.\
Needless to say that majority of mentioned papers is of purely mathematical
nature, without explicit relations to physical problems.

The combination of interval calculus with probability is tempting but certainly
uneasy to satisfy.  The papers in this area are appearing only recently\cite{FIR,Sandia,Krein2009,KreinShary,me} and still remain scarce.  Our approach,
as presented below, tries to follow this trend but seems more general and
at the same time generally easier to follow.

\section{Bridging the gap between intervals, probability and statistics}
\subsection{Interval calculus}
For some reasons, interval calculus remains still largely unknown
to the general computing public.  It is much younger than the idea of
complex numbers but it is with us already since $\sim\,$1960\cite{Ray}
and proves to be very useful, especially in computer calculations.\
In fact, it was developed first of all to put under strict control the uncertainties
of results produced by various computers operating with finite precision arithmetic,
often with different machine word lengths and/or representations of real numbers,
thus necessarily introducing rounding errors.\
Informally, interval arithmetic makes possible to compute guaranteed ranges
of mathematical expressions when exact ranges of all their components are known.
By \emph{guaranteed} we mean that the so obtained results contain true
ranges with certainty.  It doesn't mean, however, that those ranges are equal.
It happens quite often that interval result overestimates the true result -- but
never underestimates it.

For those completely unfamiliar with interval computations, we provide
a brief introduction in \ref{A} and \ref{B}.  Readers interested in more
details are referred to classical books\cite{Ray, book-intvals, Neum}
or to the nice Wikipedia page\cite{web-intvals}.
Here we only mention the notation used in the rest of this article:
symbols like $x$, $p$, or $f$ are for real variables, parameters or functions,
while their interval counterparts will be written in different font as
$\mathsf{x}$, $\mathsf{p}$, or $\mathsf{f}$, respectively.\
Greek letters always represent real quantities.

\subsection{Connection with probability and statistics}
At the first sight, there cannot be any direct connection between interval calculus
and probability or statistics.  While intervals are always guaranteed to contain the
true values then probability and statistics operate rather with imprecise quantities,
describing them in terms of most likely (expected) values and estimating their
standard deviations.  On the other hand, the very existence of well known term\
\emph{confidence interval}\ strongly suggests that such connection\ \emph{is}\
possible.

One might also ask why not to simply rewrite well known procedures, like Least Squares
Method, into their interval formulations?  This is indeed possible when experimental
uncertainties are known as guaranteed intervals (i.e. containing true values with
probability equal exactly to unity).  Doing so we are lead to system
of interval equations.  But even in the simplest case, when all equations are
linear, we encounter few serious problems.  First, we have to decide which kind
of solutions we are looking for -- as there are many possible classes to choose from.\
In what follows, we will consider only the so called \emph{united solutions} set.\
United solution set is an~interval generalization of ordinary set
of solution of a~system of equations.\  In usual arithmetics, the vector $\vec{p}$
of unknowns belongs to the solution set iff for all considered equations the following
equality holds:
\begin{equation}
L_i(\vec{p}) = R_i(\vec{p}),
\end{equation}
where $L_i(\cdot)$ and $R_i(\cdot)$ are left- and right-hand sides of equation $i$,
respectively.

However, when operating with intervals, both $L_i(\cdot)$ and $R_i(\cdot)$ are
not just real numbers, but intervals, each containing infinitely many numbers.
In this case a~subset ${\cal S}$ of searched parameters space certainly 
\emph{does not belong} to the united solutions set when
\begin{equation}
L_i({\cal S})\,\cap\,R_i({\cal S}) = \varnothing\quad {\rm for\ at\ least\ one\ } i
\end{equation}
($\varnothing$ is an empty set).
It may be somewhat unexpected, but the opposite, i.e. 
$L_i({\cal S})\,\cap\,R_i({\cal S}) \ne \varnothing$\ for all $i$\
does \emph{not} guarantee that ${\cal S}$ contains at least one true solution
of our system.\  In other words, ${\cal S}$ is only a~set of \emph{possible}
solutions.

One more comment is in order here.  It is rather unlikely that the solution set
${\cal S}$ is a~single multidimensional interval.  More often it is a~rich
composition of many `small' intervals (boxes), sometimes counted in thousands.
It is not a~comfortable situation when computer memory (or disk space) required
merely to store such a~set greatly exceeds the storage needed for original data.
Additionally, any simple operation on ${\cal S}$ becomes time-consuming task
as it has to be performed on each member of set ${\cal S}$.\  This is probably
the main reason why interval computations are still rare, even in cases when
the observed data can be considered to be \emph{guaranteed} intervals.
Of course, one might use less precise description of solution set ${\cal S}$,
say in form of intervals describing minimal and maximal values of each
parameter in turn.\
The drawback is that set ${\cal S}$ will usually occupy only a~very small part
of so defined single big box.

In further consideration we will need only one fact from probability
and statistics, namely the famous Chebyshev inequality (1874):
\begin{equation}\label{Cheb}
\Pr\,\left(\,\left|\,x^{\phantom{|}}\!-\mathrm{E}\left(x\right)\,\right|\,>\,
\xi\sigma\,\right)\,\leqslant\,\frac{1}{\xi^2}\quad\mathrm{valid\ for\ }\quad \xi> 1.
\end{equation}
It quantifies the probability of large deviations of measured value $x$
from its expected value $\mathrm{E}(x)$.\  It is valid for any probability
density function, if only $\mathrm{E}(x)$ and variance $\sigma^2$
both exist and are finite.

\section{The algorithm}
\subsection{Preliminaries}

As usually, we start with a~set of $N$ uncertain measurements
$y_1\pm\sigma_1,\,y_2\pm\sigma_2,\,\ldots,\,y_N\pm\sigma_N$, 
obtained at the corresponding values of control variables
$\mathsf{x}_1,\,\mathsf{x}_2,\,\ldots,\,\mathsf{x}_N$.\
Control variables, $\mathsf{x}$'s, are often just real numbers but
may be multidimensional entities and/or uncertain as well.\  We also
have a~model $f$, containing $k$ unknown parameters
$p_1,\,p_2,\,\ldots,\,p_k$ and relating uniquely every $y_i$ with $x_i$.\
The relation $f$ most often takes the form of algebraic equation
\begin{equation}\label{explicit}
y_i=f\left(x_i, p_1, p_2,\,\ldots,\,p_k\right),\quad\,i=1,\,\ldots,\,N
\end{equation}
(one equation for each individual measurement $y_i$, taken at
always the same, fixed set of unknown parameters
$p_1, p_2, \ldots, p_k$).

Sometimes our problem is more complicated and cannot be written
in explicit form, as in (\ref{explicit}), but rather as an implicit formula
\begin{equation}\label{implicit}
f\left(x_i, y_i, p_1, p_2,\,\ldots,\,p_k\right)=0,\quad\,i=1,\,\ldots,\,N
\end{equation}
For purely numerical reasons (see \ref{A}) it may be
desirable --- if possible --- to write relation (\ref{implicit})
in still another, but equivalent form
\begin{eqnarray}\label{twosided}
f_{\mathrm L}\left(x_i, y_i, p_1, p_2,\,\ldots,\,p_k\right)\!\!\!&\!=\!&\!\!\!%
f_{\mathrm R}\left(x_i, y_i, p_1, p_2,\,\ldots,\,p_k\right)\\
&&\ \qquad\qquad i=1,\,\ldots,\,N\nonumber,
\end{eqnarray}
where $f_{\mathrm L}$ and $f_{\mathrm R}$, treated separately, have all
their (interval) arguments appearing at most once, i.e.\ without repetitions.

From now on, our measurements will be represented as intervals:
$y_i\pm\sigma_i\,\rightarrow\,\mathsf{y}_i = \left[\,y_i-\xi\sigma_i,\
y_i+\xi\sigma_i\,\right]$, with $\xi$, called \emph{extension factor},
equal to one unless noted otherwise.\  Note that in interval calculus
the above range should be guaranteed  to contain the true value
with probability equal to exactly one.\  This requirement is satisfied
only when $\sigma_i$ is equal to maximum absolute deviation,
as specified by measuring instrument maker, and $\xi=1$.\
But even then, we may face the problem of outliers; either because
our instrument is malfunctioning or due to undetected data transmission
errors.\  In all other situations, when $\sigma_i$ is a~standard
deviation of the measurement $y_i$, even for arbitrarily large $\xi$
we have
\begin{equation}
\mathrm{Pr}\,\bigl(\,(\mathrm{true\ value\ of}\ y(x_i))\in\,%
\left[\,y_i-\xi\sigma_i,\,y_i+\xi\sigma_i\,\right]\,\bigr) < 1
\end{equation}
and the inequality is sharp.\
This observation may suggest that interval calculus is completely
unsuitable for the kind of calculations we would like to perform.\
It will be shown below that such a~view is unjustified.

\subsection{Main idea}

\begin{figure}[h]
\includegraphics[width=\columnwidth, 
angle=0, keepaspectratio]{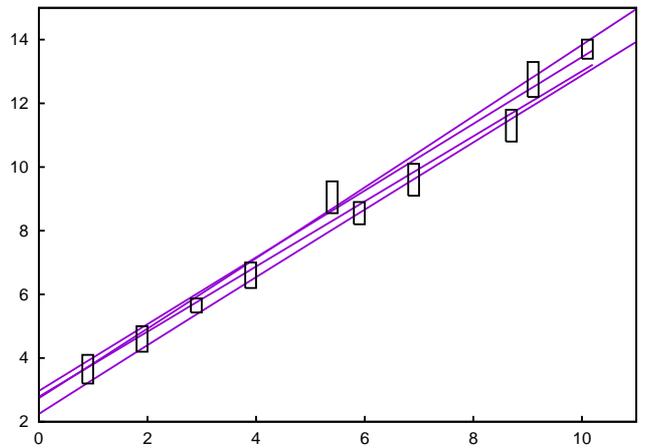}
\caption{Straight line fitting by guess. Rectangles represent results of uncertain
measurements $y_i$ taken at uncertain locations $x_i$.\  Not every trial line
is equally good: two of them miss the measurement at $x\approx{10}$, and
two others do not intercept the uncertainty rectangle near $x\approx{6}$.\
On the other hand, it is obvious that any line with negative slope (not shown)
is much worse: it will cross $0$, $1$, or at most $2$ rectangles.}
\label{straight}
\end{figure}

Our idea is illustrated in Fig.~\ref{straight}.\  When presented a~set of
uncertain measurements supposed to lay on a~straight line, we can
quickly estimate its slope and offset by simply using a~ruler.  It is rather
difficult to say which so obtained line is the `best,'\ but after few trials we
are able to estimate the sensible ranges of both relevant parameters.\
Our algorithm only formalizes those simple actions.  Its main steps are:
\begin{enumerate}\setcounter{enumi}{-1}
\item Start with the box of all fitted parameters, large enough to contain
the solution, and make it the first and only element of list $ \mathcal{L}$.
Establish \emph{unit lengths} for all searched parameters (for explanation
see the end of section~\ref{details}).
\item Pick the largest box $\mathsf{V}$ from list $\mathcal{L}$ and remove it
from list.
\item Bisect the box $\mathsf{V}$, by halving its longest edge,  to obtain
two offspring boxes, $\mathsf{V}_{\mathrm L}$ and $\mathsf{V}_{\mathrm R}$.
\item Perform admissibility tests on boxes $\mathsf{V}_{\mathrm L}$ and
$\mathsf{V}_{\mathrm R}$.  Discard box, if it appears certainly unsuitable, or
append it to the list $\mathcal{L}$ otherwise.
\item Stop when the list $\mathcal{L}$ is empty or contains only elements
being either small or certified boxes.\  Otherwise go to step~$1$.
\end{enumerate}

%\begin{center}
%\end{center}

\subsection{Details of operations}\label{details}
\begin{itemize}
\item
Bisection means halving the longest edge of the box $\mathsf{V}$.\
More precisely: if $V_{c}^{m}$ is the center of the longest edge $m$
of the original box $\mathsf{V}$ then
$\mathsf{V}_{\mathrm L}^{m}=[\underline{\mathsf{V}}^{m}, V_{c}^{m}]$~\
and~\
$\mathsf{V}_{\mathrm R}^{m}=[V_{c}^{m}, \overline{\mathsf{V}}^{m}]$,
while all the remaining components ($\ne{m}$) are exact copies
of those of parent box $\mathsf{V}$.\
This way $\mathsf{V}=\mathsf{V}_{\mathrm L}\,\cup\,\mathsf{V}_{\mathrm R}$,
what means that no point within the original search area will ever be missed
by the algorithm.
On the other hand we also have 
$\mathsf{V}_{\mathrm L}\,\cap\,\mathsf{V}_{\mathrm R}\ne\varnothing$,
since offspring boxes always share a~common face.\  We will need this
feature at later stages.

\item
Testing means counting `hits.'\   By `hit' we understand the event\
$\mathsf{f}\left(\mathsf{x}_i, \mathsf{V}\right)\,\cap\,\mathsf{y}_{i}\,\ne\,\varnothing$~\
(compare with formula (\ref{explicit})), or non-empty intersection of
$\mathsf{f}_{\mathrm L}\left(\mathsf{x}_i, \mathsf{y}_i, \mathsf{V}\right)$
and
$\mathsf{f}_{\mathrm R}\left(\mathsf{x}_i, \mathsf{y}_i, \mathsf{V}\right)$
--- when formula (\ref{twosided}) is at work.\  Box $\mathsf{V}$ should
pass the test, when number of hits exceeds number of misses (empty
intersections).  But we shouldn't ignore constraints, if there are any.
Violating of at least one constraint immediately invalidates the box,
if only this violation is certain.

For example, if we require two unknown parameters $p_m$ and $p_n$ to be
equal, then the investigated box $\mathsf{V}$ should be discarded only
when the intersection of its corresponding components is empty:
$\mathsf{V}^{m}\cap\mathsf{V}^{n}=\varnothing$.\
Non-empty intersection means that our constraint has a~chance to be
satisfied in current box and therefore $\mathsf {V}$ should be retained
for further investigations (if there is no other certainly violated constraint
within this box, of course).

It may happen, when the task is to satisfy formula (\ref{explicit}), that
in given box $\mathsf{V}$ the following inclusion occurs:
$\mathsf{f}\left(\mathsf{x}_i, \mathsf{V}\right) \subseteq \mathsf{y}_i$,
for $i=1, 2, \ldots, N$,~\ whenever
$\mathsf{f}\left(\mathsf{x}_i, \mathsf{V}\right)\,\cap\,\mathsf{y}_{i}\,\ne\,\varnothing$.
Such a~box may be safely called \emph{certified} as it needs not to be
bisected further.\  This is because any subset of $\mathsf{V}$ also
satisfies this inclusion.  It is therefore a~good idea to put such box
aside and never test it again.

\item
\emph{Small boxes} are those with diameter not exceeding
unity.\  But how can we compare searched parameters of different nature,
expressed in various units, like meters, degrees or seconds?  For this we
need to arbitrarily establish \emph{unit lengths}, individually for each
searched parameter.\  This way the lengths of all edges of our boxes
will become dimensionless numbers. Adopted unit lengths should not
exceed accuracies we expect to get, but making them too small will
result in significant 	increase of computation time.
\end{itemize}

\subsection{What next?}
\subsubsection{$\mathcal{L}$ is empty}
We are done, but certainly not satisfied, when $\mathcal{L}$
is empty.\
What could be the reason not to obtain any result at all?
Apart from obvious mistake of processing data obtained
from different experiment, or mistakenly searching parameters
outside their true ranges, we can think about the validity of
our formula~$f$.  Maybe our model $f$ is simply too rough
and is therefore unable to replicate observed features?
Maybe it is only applicable within some range of control
parameters and not outside it?

Less obvious reason for emptiness of the list $\mathcal{L}$ is
perfectly adequate model evaluated on \emph{too precise} data.\
By \emph{too precise} data we mean those with grossly underestimated
uncertainties, including the case when they are being presented
as equal to zero to the algorithm.\
It is evident that cheating doesn't pay.

Yet, the case of \emph{too precise} data need not to be completely
at lost.\  It is possible to get $\mathcal{L}\ne\varnothing$ in another
run, with significantly enlarged unit lengths.\  Of course, the standard
deviations of so obtained results may be very disappointing.  This
is the price for poor quality/inconsistent measurements.

When none of the above mentioned cases applies and the list
$\mathcal{L}$ in nevertheless empty, then we can conclude
that our model $f$ is \emph{certainly} inadequate to the
problem under investigations.

\subsubsection{$\mathcal{L}$ is non-empty}
In this case we should check whether the convex hull of all boxes
has no common parts with any face of the initial search domain.\
The presence of some boxes at the original boundary usually means
that either the initial search domain was too small (not covering
all solutions) or the corresponding unit length was selected too large.\
The second possibility will certainly occur in `pathological' cases, when
we don't want to evaluate some parameter(s) and therefore deliberately
and forcibly fix their values by setting widths of their search intervals
to zero.\
There exist still other possibilities, to be discussed later, but in any
case the algorithm should issue a~detailed warning after
encountering such a~situation.

So,  there is at least one box present on the list $\mathcal{L}$.\
Yet even a~single box contains infinite number of solutions,
what is in sharp contrast with results delivered by other point-type
routines, Monte Carlo investigations, or even population-based
approaches, like genetic algorithms.\  By the way, the list $\mathcal{L}$
with exactly one member will be an exception rather than the rule.\
More often $\mathcal{L}$ will consist of much more boxes, perhaps
counted in thousands.\  How should we report our results?

\subsection{Reporting results}
Well, first thing is to check how many solutions were found.\
The obtained boxes need not to make a~simply connected set, they
may form few disjoint clusters.\  Is it possible?  Yes, think about
fitting two non-overlapping spectral peaks (their positions, amplitudes
and half-widths) located on noisy background.\ Without constraints
we will get \emph{two} solutions, showing exactly the same two peaks
but in different order.

For this reason, the next step should be to recover the individual
simply connected components, that is clusters of neighboring boxes.\
It is the place where the property 
$\mathsf{V}_{\mathrm L}\,\cap\,\mathsf{V}_{\mathrm R}\ne\varnothing$
will be exploited extensively.\  Indeed, this part of algorithm often appears
the most time-consuming one.\ Only after this step is completed, it is
possible to process/report each solution, one after another.

\subsubsection{Original uncertainties are guaranteed limits}

In this case the final processing is rather simple.  All we have to do is
to compute the convex hull of each cluster.\  This way the guaranteed
lower and upper limit of each searched parameter are determined,
in full concordance with usual rules of interval computations.\
Final report will typically include extremal values for each parameter
as well as centers of those intervals.

\begin{figure}[h]
\includegraphics[width=0.70\columnwidth,angle=-90.0, keepaspectratio]{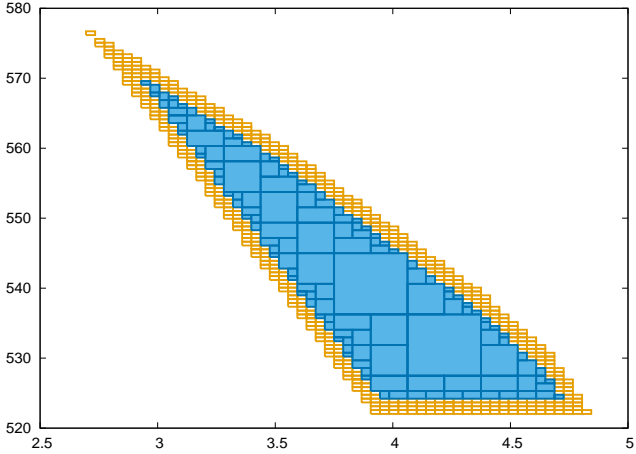}
\caption{Exemplary contents of list $\mathcal{L}$ for case of two fitted  parameters.\
Shaded boxes are \emph{certified}, thus not necessarily \emph{small}.
}
\label{wrapping}
\end{figure}

One can think about the hulls of certified boxes making each cluster.\
Looks like that by doing so, we may get `more accurate' (tight)
estimates of searched parameters, see Fig.~\ref{wrapping}.\
Unfortunately, this is a~bad idea.\  First, the cluster might contain
no certified boxes at all!  This will almost certainly happen whenever
adopted unit lengths are too large. Secondly, we will loose the rigor
of interval computations.

%\begin{center}
%\end{center}

In some applications it is essential to know guaranteed tolerances
of searched parameters. If so, then we need as a solution the `biggest'
box covering certified boxes, and only those boxes.\  However, the
solution having this property is not unique --- it depends on what the word
`biggest' means in every particular case.  In practice, some parameters
may be easy to control, while others only with excessive cost,
and so on.\  Thus it may be a~matter of user preferences which
solution is preferable.\  The case of linear equations was extensively
studied by Shary\cite{Shary}, and similar problems -- mostly related
to robotics -- were presented at numerous conferences.
Nevertheless, this topic is is out of scope of the current paper.

\subsubsection{Original uncertainties and standard deviations}
At first sight, the rigor of interval computations becomes doubtful,
when we operate on data expressed in familiar form, i.e.\ as a pair:
measurement result, $y$,  and its standard deviation, $\sigma_y$.
This is because no interval of type $\left[y-\xi\sigma_y, y+\xi\sigma_y\right]$
guarantees that the true value, $\tilde{y}$, is located within these limits,
no matter how large (but finite) is the positive extension factor~$\xi$.\
In some cases we can find the exact value of probability of such event,
most notably when the measured quantity, $y$, follows normal (Gaussian)
distribution, or --- generally --- when the probability distribution is known
(preferably in analytical form).\
In all other cases we can use the already mentioned Chebyshev
inequality (\ref{Cheb}) to rigorously estimate probabilities of interest.

\vskip 1ex
Before we proceed further, let us explain our point of view on data
fitting process, to our best knowledge never presented before.\
In short: data fitting process is like a~final step of any ordinary
(direct) measurement.\
And here is why.\  During direct measurements, the measuring instrument,
say a ruler, seems to deliver immediate answer how long is the investigated
object.\  In reality things are slightly more complicated, even in simple
cases like that one.\  Here we have a~light, which is reflected, both from
our instrument and from the object under study, and which finally reaches our eye.\
Still later on, our impression of reality is transferred to our brain,
which decides the final outcome of the measurement.\  This complicated
process may be formally described as a~superposition of several
transformations.\  So, between the input signal(s) and the final numerical
outcome(s) there are intermediate steps, some of them performed inside
measuring instrument.\  Generally, all what this machinery is doing is
selection of a~single number from the real line and presenting it as the
final result.\  Such an action is usually repeated several times, what makes
possible to estimate the most probable value of measured quantity and
its standard deviation.

This is exactly what our algorithm is doing, with the small exception
that its `measurements' are repeated infinitely many times.\  Thanks to this
observation we can think about every point within the obtained cluster
of boxes as being the result of a~single measurement -- why not?
This way our algorithm becomes a~last part in chain of transformations
normally performed by measuring instrument.\  What remains, usually
the experimenter's task (and her computer, perhaps), is to derive simplest
statistical properties of the bunch of measurements.\  But now finding
expected values of searched parameters, their standard deviations, and
correlation coefficients as well, is a~next to trivial task.  Best of all, it can be
done without any tricks, or simplifications, just by following appropriate
definitions.

In conclusion: for most popular types of measurements we are able
to find and present not only the extremal values of fitted parameters but
also highly desired, reliable estimates of their standard deviations and
correlation coefficients, for every separate cluster in turn.

\vskip 1ex
\noindent
\emph{Final note.}  It is tempting to treat on unequal basis the boxes
differing by number of hits.\  Assigning higher weights to points
located in boxes with higher number of hits will certainly result in smaller
values of standard deviations of fitted parameters --  but is it well justified?
At this moment this remains an open question.

\section{The meaning of extension factor}
Before the algorithm starts, it needs to know its input data in interval
form.\  This is easy when input data are known within guaranteed limits.\
No extension factor is needed then, correct intervals are already known.

Let's discuss all other cases now. As already mentioned, no value of
extension factor $\xi$ makes certain that true value is covered by so created
interval.  Assume that we know nothing about the distribution(s) ruling
our measurements, except that its average value and variance do exist
and are both finite.  Our goal is to find such values of unknown parameters that
resulting curve (or a~hyper-surface in multidimensional case) hits more
than half of our uncertain measurements.  Suppose such set of parameters
indeed makes sense (exists).  If so, then hits are binomially distributed, with
probability $p$ of success in a~single trial equal to the probability of true
value being located within the inspected interval.\  One might think, that all what
is needed to hit more than half of measurements is to set
$(\xi^2=2)\,\equiv\,(\xi=\sqrt{2})$ in Chebyshev inequality.\
Unfortunately, this guess is correct only for a~single measurement 
or for infinitely many measurements.  In all other cases
we need to find smallest\  $\,0 < p < 1$ satisfying inequality:
\begin{equation}\label{Poiss}
\sum_{k=0}^{\lfloor\,(N+1)/2\,\rfloor -1} {{N}\choose{k}}\,p^k\,q^{N-k}\, \leqslant
\sum_{k=\lfloor\,(N+1)/2\,\rfloor}^{N} {{N}\choose{k}}\,p^k\,q^{N-k}
\end{equation}
where $N$  is the number of uncertain data points, $q=1-p$,
and $\lfloor\,\cdot\,\rfloor$
means integer part of the argument.\  From Chebyshev inequality (\ref{Cheb}) we
immediately have $1-p \leqslant 1/\xi^2$ and thus the minimal value
of extension factor $\xi =1/\sqrt{1-p}$.

\begin{figure}
\includegraphics[width=0.7\columnwidth,angle=270.0, keepaspectratio]{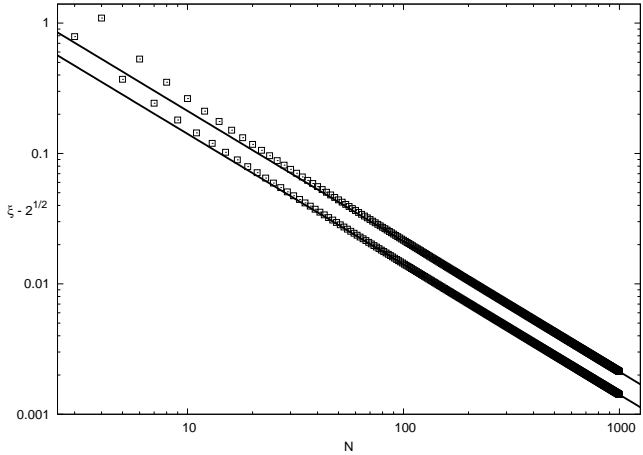}
\caption{Extension factor, $\xi$\ (decreased by $\sqrt{2}$), \emph{vs.}
the number of fitted data entities, $3\leqslant{N}\leqslant{1000}$.\
Two distinct subsequences are clearly visible, both well approximated
by straight lines.\ Note the logarithmic scale in both coordinates.
}
\label{sequence}
\end{figure}

In order to satisfy the inequality~(\ref{Poiss}), one has to resort
to numerical calculations.\  The results, for $N\leqslant{30}$, are
presented in Table~\ref{tabelka}, together with minimal required probability
of success (hit) in single attempt ($p=\,$min.~Pr).\  Somewhat unexpectedly,
the sequence $\left\{\xi_N\right\}$ appears to consist of two distinct
subsequences: one for even and the other for odd $N$, as illustrated in
Fig.~\ref{sequence}.\  Both subsequences are decreasing and
converge to the same limit: $\lim_{N\rightarrow\infty}\,\xi_N=\sqrt{2}$.\
For practical purposes extension factors may be approximated by
\begin{eqnarray}
\xi_N&\approx&\sqrt{2}\left(1+\frac{3}{2N}\right)\qquad\mathrm{for~}\ N\ \mathrm{even}\\
\xi_N&\approx&\sqrt{2}\left(1+\frac{1}{N}\right)~\ \qquad\mathrm{for~}\ N\ \mathrm{odd}
\end{eqnarray}
Both approximations are from below, with the ratio of true extension factor
and its approximated value dropping below $1.0010$ for $N\geqslant{35}$ ($N$ odd)
or $N\geqslant{54}$ ($N$ even).

It is worth noticing that for $N\in[2,6]$ (see Table~\ref{tabelka}) we should
use \mbox{$\xi>1$}, i.e.\ we should extend original intervals, even for
normally distributed uncertainties.\  One might think that for normally
distributed data it should make a~sense to shrink their intervals
(apply\  $\xi<1$) whenever
$N\geqslant{7}$.\  Unfortunately, this is not recommended, as in such
case we are loosing the solid ground of Chebyshev inequality~(\ref{Cheb}).\

In effect, to avoid this unreasonable temptation, we leave  the last
column of Table~\ref{tabelka} mostly unfilled.

Note also the unexpected relation
\begin{equation}
\Pr\,(N-1) > \Pr\,(N) < \Pr\,(N+1)
\end{equation}
valid for odd $N$.\  Analogous inequality is satisfied by~$\xi$, what
suggests potentially more tight estimates for data sets containing
odd number of measurements --- at least when using the proposed
approach.\  In fact, we are dealing here with the old truth: odd number
of voters will never generate a~tie what might happen when
the number of voters is even.\
Another surprising observation is that the general idea
\emph{more is better} indeed works, but only for $N\,\geqslant\,4$.

\vskip 1ex
\emph{Final remark:}  Some measurements deliver not just
a~single number but rather few components at once, say two or
three components of a~vector.\  The extension factor, $\xi$, should be
modified in such situations accordingly.  Namely, it should be replaced
with $\xi\leftarrow\,\xi^\frac{1}{D}$, where $D$ is the number
of individual components making single measurement.\  Not doing so will
result in needlessly overestimated uncertainties of fitted parameters.

\begin{table}[ht]
\caption{Extension factor $\xi$ \emph{vs.} the number of collected
measurements $N$, valid for unknown distribution of their uncertainties.\
$N_{\mathrm{min}}=1+\lfloor\,N/2\,\rfloor$ is the required minimal
number of hits.  The column marked as `min.~Pr'  shows minimal
probability of a single measurement to guarantee that $\mathrm{Pr}%
\left(N_\mathrm{hit}\geqslant{N}_\mathrm{min}\right)>1/2$.\
The last column shows extension factors for normally distributed
measurements.\  Values below unity are not shown.}
\begin{center}
\begin{tabular}{|rrrrr|}
\hline
$N$ & $N_{\mathrm{min}}$ & $\xi$~unknown & min.~Pr&$\xi$~Gauss\\
\hline\hline
     1  &   1  &  $ \sqrt{2}$ & 0.500000 &                 \\ %1.414214  
     2  &   2  &   1.847759   & 0.707107 & 1.051801 \\
     3  &   2  &   2.201664   & 0.793701 & 1.263802 \\ 
     4  &   3  &   2.507033   & 0.840896 & 1.408092 \\ 
     5  &   3  &   1.785116   & 0.686190 & 1.007259 \\
     6  &   4  &   1.944591   & 0.735550 & 1.115935 \\
     7  &   4  &   1.657220   & 0.635884 & \\ 
     8  &   5  &   1.766335   & 0.679481 &\\
     9  &   5  &   1.594986   & 0.606915 &\\
    10 &   6  &   1.678127   & 0.644900 &\\
    11 &   6  &   1.558150   & 0.588110  &\\
    12 &   7  &   1.625364   & 0.621471 &\\
    13 &   7  &   1.533792   & 0.574923 &\\
    14 &   8  &   1.590223   & 0.604557 &\\
    15 &   8  &   1.516488   & 0.565167 &\\
    16 &   9  &   1.565126   & 0.591773 &\\
    17 &   9  &   1.503560   & 0.557658 &\\
    18 & 10  &   1.546302   & 0.581774 &\\
    19 & 10  &   1.493535   & 0.551699 &\\
    20 & 11  &   1.531658   & 0.573738 &\\
    21 & 11  &   1.485532   & 0.546856 &\\
    22 & 12  &   1.519939   & 0.567140 &\\
    23 & 12  &   1.478997   & 0.542843 &\\
    24 & 13  &   1.510348   & 0.561625 &\\
    25 & 13  &   1.473559   & 0.539463 &\\
    26 & 14  &   1.502354   & 0.556947 &\\
    27 & 14  &   1.468964   & 0.536577 &\\
    28 & 15  &   1.495588   & 0.552929 &\\
    29 & 15  &   1.465029   & 0.534084 &\\
    30 & 16  &   1.489787   & 0.549441 &\\
%  999 & 500&   1.415631   & 0.501001 &\\
% 1000& 501&   1.416339   & 0.501500 &\\
\hline    
\end{tabular}\label{ext}
\label{tabelka}
\end{center}
\end{table}

\section{An example}
\begin{figure}
\includegraphics[width=0.7\columnwidth,angle=270.0, keepaspectratio]{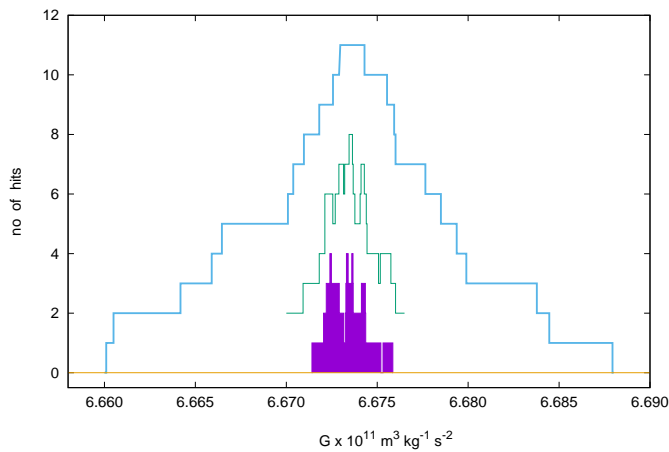}
\caption{Coverage of the search domain with original measurements
(shaded part), with data modified by extension factor $\xi=1.558150$
(shifted up by $2$~hits for clarity) -- as appropriate for $N=11$ measurements
of unknown origin, and following methodology of CODATA 2010
report, i.e.\ with $\xi=14$.
}
\label{G-covering}
\end{figure}

As an example we present how our algorithm deals with
experimental data on Newton gravitational constant $G$.\  This fundamental
physical constant still remains the least precisely known\cite{troublesomeG}.\ 
The 11 original measurements, presented in CODATA report\cite{CODATA}
and not repeated here,  are not quite consistent, as can be seen from
shaded area in Fig.~\ref{G-covering}.\  No more than four of them
overlap anywhere within the full range, with small interval
$[6.67520,\,6.67532]\,\times{10}^{-11}\,$m$^3$kg$^{-1}$s$^{-2}$
being not covered at all, and another one,
$[6.67315,\,6.67324]\,\times{10}^{-11}\,$m$^3$kg$^{-1}$s$^{-2}$,
near the center of `solid body,' containing just one measurement.\
The CODATA committee found no good reason to discard any of those
$11$~measurements, so they decided to extend all uncertainties by the
factor 14 before proceeding.\  The resulting bell-shaped `coverage curve,'
which may be considered to be proportional to the probability density
function, is also shown in Fig.~\ref{G-covering}.\ It remains unclear what
were the other steps of procedure, but the final result, the recommended
value of Newton gravitational constant, is:
$G = 6.67384(80)\times{10}^{-11}\,$m$^3\,$kg$^{-1}\,$s$^{-2}$.

We find only the narrow range
$[6.673449,\,6.673636]\,\times{10}^{-11}\,$m$^3\,$kg$^{-1}$s$^{-2}$
being covered by minimally required majority of (expanded) measurements
($6$ out of $11$).\  The result produced by our algorithm is:
$G = 6.673542(54)\times{10}^{-11}\,$m$^3\,$kg$^{-1}\,$s$^{-2}$.\
Not surprisingly our result is roughly 14 times more precise.

\section{Advantages and disadvantages}
Advantages:
\begin{itemize}
\item
recovering unknown parameters from implicit dependencies
is equally easy as from explicit formulas
\item
robustness against outliers, up to $50\,$\%
\item
obtained uncertainties (variances) are never underestimated
\item
no need to apply any `error propagation law,' often questionable
\item
high flexibility.  The same algorithm may be used for detecting
outliers, hypotheses testing, solving systems of nonlinear
equations, possibly containing uncertain parameters,  or just for
simulations of complex, implicit models with uncertain parameters.\
The presented approach may be also useful in metrology, for
inter-laboratory comparisons.
\item uncertainties in both coordinates do not pose any problem
and are handled naturally, including non-linear and implicit cases.\ 
In fact, we have found in literature only two articles\cite{xy,Krystek}
describing straight line fitting with uncertainties in both variables;
general non-linear cases seem not to be discussed at all.
\end{itemize}

Disadvantages:
\begin{itemize}
\item
worst case complexity is exponential in the number of fitted parameters, thus
\item
impractical when the number of unknowns is large
\item
the computed standard deviations may be overestimated by unknown factor.
\end{itemize}

\section{Conclusions}
It is hoped that the presented algorithm will soon replace
a~great deal of existing optimization procedures.  In author's opinion,
interval computations deserve to become soon as familiar
to experimentalists as are complex numbers to electrical engineers.

\appendix
\section{Brief introduction to interval computations}
\label{A}
\subsection{Some definitions and important properties}
An \emph{interval}\  ${\mathsf x}$ is a compact and finite subset of
a~real axis:
$${\mathsf x} = \left\{\,\mbox{\mm R}^{\phantom{!}}\!\ni x:\,
\underline{\mathsf x} \leqslant x \leqslant \overline{\mathsf x}\,
\right\}\, \mathop{=}^{\mathrm{def}}\,
\left[\,\underline{\mathsf x},\,\overline{\mathsf x}\,\right]\,\subset\,\mbox{\mm R},$$
where both $\underline{\mathsf x}$ and $\overline{\mathsf x}$ are finite.\
It may be thought to be a~representation of a~real number, certainly
located somewhere between
$\underline{\mathsf x}$ and $\overline{\mathsf x}$, inclusive,
but unknown otherwise.\  A special case is
$\mathsf{x} = \left[\,a,\,a\,\right]$ (a.k.a.\ \emph{thin interval} or
\emph{singleton}), identified with the real number $a$.
The set of all intervals is usually denoted as\ $\mbox{\mm IR}$.

It is easy to define interval counterparts of ordinary arithmetic operations:
\begin{eqnarray}
\mathsf{x}+\mathsf{y}&=& \left[\,\underline{\mathsf{x}}+\underline{\mathsf{y}},\,
\overline{\mathsf{x}}+\overline{\mathsf{y}}\,\right]\\
\mathsf{x}-\mathsf{y}&=& \left[\,\underline{\mathsf{x}}-\overline{\mathsf{y}},\,
\overline{\mathsf{x}}-\underline{\mathsf{y}}\,\right]%
\qquad\left(\,\mathsf{a}\!\ne\!0\,\Rightarrow\,\mathsf{a}\!-\!\mathsf{a}\!\ne\!0\ %
\right) \\
\mathsf{x}\,\cdot\,\mathsf{y}&=&\left[\,\min\,{\mathcal Z},\, \max\,{\mathcal Z}\,\right]
\end{eqnarray}
where ${\mathcal Z}$ is a~four--element set:~\ 
$\mathcal{Z}= \left\{\,
\underline{\mathsf{x}}\,\underline{\mathsf{y}},\,
\underline{\mathsf{x}}\,\overline{\mathsf{y}},\,
\overline{\mathsf{x}}\,\underline{\mathsf{y}},\,
\overline{\mathsf{x}}\,\overline{\mathsf{y}}\,\right\}$.
Division is defined, for $\mathsf{y}\,\notni\,0$\ as:~\
$\mathsf{x}\,/\,\mathsf{y} = \mathsf{x}\cdot{1/\mathsf{y}},$~\
where~\
$1/\mathsf{y}=
\left[\,1/\overline{\mathsf y},\,1/\underline{\mathsf y}\,\right]$,~\
and remains undefined otherwise (as usually).\  It may be checked
that so defined arithmetic operations produce all possible results
of $x\,\boxdot\,y$\ for any pair $(x,y)$ satisfying $x\,\in\,\mathsf{x}$
and\ $y\,\in\,\mathsf{y}$,
and \emph{only} those results (here $\boxdot$ stands for any of
$+$, $-$, $\cdot$ or $/$).\  However, more complicated arithmetic
expressions may happen to overestimate the true range.\ Specifically,
we generally have:~\ $\mathsf{x}\,\left(\mathsf{y}+\mathsf{z}\right)
\subseteq\, \mathsf{x}\,\mathsf{y}+ \mathsf{x}\,\mathsf{z}$\
~for~\ $\mathsf{x, y, z}\,\in\,\mbox{\mm{IR}}$, not the equality.\ 
Nevertheless, the following theorem holds\cite{Ray}:
\vskip 1 ex
Theorem (\emph{Fundamental Theorem of Interval Arithmetic})\\
Let $f(x_1, x_2 , \ldots, x_n)$ be an explicitly defined real function. Then
evaluating $f$  `in interval mode' over any interval inputs $\left(%
\mathsf{x}_1, \mathsf{x}_2, \ldots, \mathsf{x}_n\right)$ is
guaranteed to give a set $\mathsf{f}$ that contains the range of $f$
over those inputs.
\vskip 1 ex
The above theorem is true, but in practice we often obtain overestimated
results, i.e.\ intervals wider than necessary.\  To avoid such undesired
situations, we should --- whenever possible --- write complex interval
expressions in form with each interval variable appearing exactly once.\
For example, to compute the resistance~\ $R$ of two resistors $R_1$
and $R_2$, connected in parallel, we normally use the formula~\
$R=R_1\cdot{R}_2/(R_1+R_2)$.\
When $R_1$ and $R_2$ are uncertain, it is better to compute
their equivalent resistance as~\ $\mathsf{R}=\left(1/\mathsf{R}_1 +
1/\mathsf{R}_2\right)^{-1}$.

There is another subtlety, not mentioned until now.\  Our algorithm
extensively exploits the `obvious' property:
$\mathsf{x} \subset \mathsf{y} \Longrightarrow
\mathsf{f}(\mathsf{x}) \subseteq  \mathsf{f}(\mathsf{y})$,
which need not to be true.\ Functions $\mathsf{f}$ satisfying this
relations are called \emph{monotonously inclusive}.  At this place
it is enough to say that all ordinary (`calculator') functions have this
property.  Nevertheless exceptions sometimes happen and among
the suspected functions are those containing min and/or max.

When dealing with interval computations on a~computer, that is
with finite precision, it is also very important to properly round all
the intermediate results, as well as the final one.\  Proper rounding
means outwards rounding, i.e.\ lower (left) endpoint has to be rounded
towards minus infinity, while the other one --- towards plus infinity.
Fortunately, the existing interval software packages have this feature
built in. Sometimes, however, it is highly recommended to perform
such an action explicitly.

Interval $n$--dimensional vectors, i.e.\ objects belonging to
Cartesian product
$\mbox{\mm IR}^n=\mbox{\mm IR}\times\mbox{\mm IR}%
\times\ldots\times\mbox{\mm IR}$, are often called
\emph{boxes}, for obvious reasons.\  We will need to know
how large are our boxes.  The box's \emph{diameter} is a~real
number, defined as the length of its longest edge:
\begin{equation}\label{diam}
\mathrm{diam}\,\left(\,\mathsf{x}_1,\,\ldots\,\mathsf{x}_{\mathrm{N}}\,\right)%
= \max\,\left(\,\overline{x}_1-\underline{x}_1,\,\ldots\,,\,%
\overline{x}_{\mathrm{N}}-\underline{x}_{\mathrm{N}}\,\right)
\end{equation}

\section{Set theory operations on intervals}\label{B}
Intervals are sets and therefore also the set--theory operations may
be performed on them.  Here we sketch only two:
\begin{itemize}
\item intersection
\begin{equation}
\mathsf{a}\,\cap\,\mathsf{b} = \left[\max\,\left(\underline{\mathsf{a}},
\underline{\mathsf{b}}\right),\,
\min\,\left(\overline{\mathsf{a}},\overline{\mathsf{b}}\right)\right]
\end{equation}
When $\mathsf{a}$ and $\mathsf{b}$ happen to be disjoint,
then the above formula will necessarily produce illegal
result, not an element of $\mbox{\mm IR}$, i.e.\ the one with left
endpoint value higher than right endpoint.\ If this is
the case, then we should replace the so obtained result with an
\emph{empty interval}, see below.

\item convex hull
\begin{equation}
\mathsf{hull} \left(\mathsf{a},\mathsf{b}\right) = 
\left[\,\min\,\left(\underline{\mathsf{a}},\,\underline{\mathsf{b}}\right),\,
\max\,\left(\overline{\mathsf{a}},\overline{\mathsf{b}}\right) \right]
\end{equation}
This operation is an interval counterpart of union of two sets,
with result being again an element of $\mbox{\mm IR}$.\
We always have $\mathsf{a}\cup\mathsf{b}\subseteq\,
\mathsf{hull} \left(\mathsf{a},\mathsf{b}\right)$, with equality
occurring only for arguments having non--empty intersection.
Therefore we can say that interval $\mathsf{hull}$ possibly
overestimates ordinary union of $\mathsf{a}$ and $\mathsf{b}$.

\item empty interval\\
It is easy to see that arbitrary illegal interval $\mathsf{a}=%
\left[\underline{\mathsf{a}}, \overline{\mathsf{a}}\right]$\
with $\underline{\mathsf{a}} > \overline{\mathsf{a}}$,
doesn't guarantee the satisfaction of the otherwise obvious property
$\mathsf{a}\,\cup\,\mathsf{b} = \varnothing\,\cup\,\mathsf{b} =%
\mathsf{b}$, as one might expect.\
Therefore we need to define an empty interval in a~special form, the one
making possible to always obtain correct results during computer
calculations.\  The suitable choice is
\begin{equation}
\varnothing = [\,HUGE,\,-HUGE\,],
\end{equation}
where $HUGE>0$ is the largest machine number.
\end{itemize}

\section*{Acknowledgments}

Current author's interest in interval computations started after Ramon E.~Moore
published the paper\cite{seminal}  (1977) showing the power of interval
Newton method applied to nonlinear equations.  Author is deeply 
indebted to him as well as to R.~Baker Kearfott, Vladik Kreinovich, Sergey P.~Shary
and Sergei I.~Zhilin for words of encouragement and occasional e-mail discussions
on interval-related problems. 

For long time the interval
computations were developed with no visible link to physical problems, until the
largely overlooked conference presentation\cite{SIAM2002} given in 2002.\
This article presents a~working algorithm following the ideas sketched there.

This work has been done as a~part of author's regular duties in the Institute
of Physics, Polsh Academy of Sciences, and was not funded otherwise.

\section*{References}

%% References
%%   \citet{key}         ==>   Author [#]


\bibliographystyle{model6-num-names}
\bibliography{<your-bib-database>}

%% not want to use model6-num-names.bst.


\end{document}